\documentclass[12pt]{article}
\usepackage{graphics}
\usepackage{graphicx}
\DeclareGraphicsExtensions{.pdf}
\usepackage{float} 
\usepackage{amsmath}
\usepackage{slashed}
\usepackage{amsfonts}   
\usepackage{amssymb}

\usepackage{color}

\begin{document}
\date{}
\title{{\bf{\Large Metallic transports from accelerating black holes }}}
\author{
 {\bf {\normalsize Dibakar Roychowdhury}$
$\thanks{E-mail: dibakar.roychowdhury@ph.iitr.ac.in}}\\
 {\normalsize  Department of Physics, Indian Institute of Technology Roorkee,}\\
  {\normalsize Roorkee 247667, Uttarakhand, India}
\\[0.3cm]
}

\maketitle

\begin{abstract}
We probe four dimensional accelerating black holes with D-brane and build up the notion of metallic holography for spacetime with negative cosmological constant. We explore various thermodynamic entities associated with the boundary QFT at low temperatures and finite chemical potential. The DC conductivity in the boundary QFT is enhanced due to the effects of black hole acceleration in the bulk counterpart. We further compute resistivity in different temperature regime, which reveals a new quantum liquid phase with dynamic critical exponent $ z=3 $.
\end{abstract}
\section{Introduction and Overview}
The $ C $-metric, as it was originally proposed by Levi-Civita \cite{levi} and subsequently extended by others \cite{Plebanski:1976gy}-\cite{Hong:2003gx}, has gained renewed attention in the recent years due to its several remarkable features. The metric essentially corresponds to an axisymmetric vacuum solution of Einstein's equations representing an accelerating black hole in four spacetime dimensions.

Understanding various thermodynamic aspects and an interpretation of Bekenstein-Hawking area law for accelerating black holes has been a challenge for decades. This was addressed until very recently in a series of papers\footnote{See \cite{Gregory:2017ogk}-\cite{Appels:2018jcs}, for a nice and comprehensive set of reviews on the subject.} \cite{Appels:2016uha}-\cite{Anabalon:2018qfv}. Following their analysis, a number of other promising directions had opened up which include - accelerating black holes in ($2+1$) dimensions \cite{Arenas-Henriquez:2022www}-\cite{Arenas-Henriquez:2023hur},  supersymmetric accelerating $AdS_4$ black holes \cite{Cassani:2021dwa}, holographic and thermodynamic aspects of charged accelerating black holes \cite{Kubiznak:2024ijq}-\cite{Jiang:2021pzf}.

Given the state of the art, the purpose of the present paper is to probe the boundary QFT (dual to accelerating black holes) at finite density and/or chemical potential and to build up the notion of metallic holography following the lines of \cite{Hartnoll:2009ns}-\cite{Lee:2010ii}. In order to address this question in detail, we ``probe'' accelerating black hole background by D-brane, where we always work with the slow acceleration ($A \ell \ll 1$) limit of the black hole and consider no mutual interaction between the D-brane and the cosmic string in the bulk. 

\begin{figure}
\begin{center}
\includegraphics[scale=1.2]{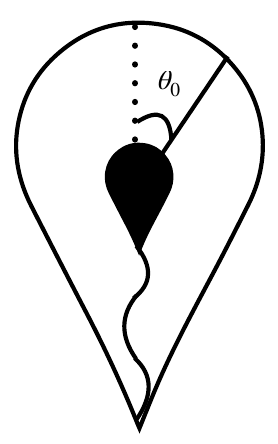}
  \caption{We show the Poincare disc for an accelerating black hole spacetime. The black hole is suspended by means of a cosmic string that corresponds to a conical deficit at the south pole of the two sphere. The dotted arc shows the new location of the ``modified'' AdS boundary. The D-brane lives on $ \theta= \theta_0 $ hyper-plane, where one of the world-volume directions lies between the horizon $ (r_+) $ and the modified AdS boundary $r_b=-\frac{1}{A \cos \theta_0}$.} \label{fig}
  \end{center}
\end{figure}

In order to compute various observables in the boundary theory, we choose $ \theta= \theta_0$ and study the dynamics of D-brane in this subspace (Fig.\ref{fig}). The corresponding dual QFT lives at a distance $ r_b=-\frac{1}{A\cos\theta_0} $ from the centre of the bulk spacetime. Using holographic techniques, we define grand-canonical partition function and estimate various thermodynamics entities as well as linear response functions for the boundary QFT at finite density and temperature. The computation of thermodynamic entities and charge transports are based on the assumption that the partition function can be defined considering a quasi-static equilibrium. This assumption is valid only in the domain of infinitesimal acceleration ($ A \ll 1 $), where the external force due to cosmic string could be thought as a perturbation over the equilibrium (grand-canonical) partition function.

We compute DC charge current in the boundary QFT which is enhanced due to the acceleration of the black hole. Here the cosmic string acts as an additional driving agency on top of the external electric field ($ \mathcal{E} $) due to probe D-branes. Therefore, to summarise, the boundary current receives contributions both due to the physical acceleration of the D-brane (or the black hole) and the external electric field ($ \mathcal{E} $). By Ohmic conductivity ($ \sigma_b $) at the boundary QFT, we always refer to the current that is produced due to $ \mathcal{E} $.

The acceleration of the black hole/D-brane causes a uniform current through the spacetime even in the absence of an external electric field ($ \mathcal{E} $). In other words, a constant acceleration ($A$) in bulk corresponds to a uniform ``driving force'' that is applied on the charge carriers in the boundary QFT and thereby producing a steady background current. 

These charge carriers are of two types- (i) one that is categorised as thermally excited charge pairs and (ii) the $ U(1) $ carriers that are added externally due to probe D-branes. It turns out that at higher temperatures, the current in the boundary QFT is dominated mostly by the thermally excited charge pairs. The $ U(1) $ carriers, on the other hand, play significant role at relatively low temperatures. One could think of these charge carriers as (quasi-particle) excitations above QFT ground state, that are drifted through the medium in the presence of external driving agencies. The medium also offers a ``drag force'' to these charge particles which eventually balances these external forces.  

The organisation for the rest of the paper is as follows. We begin with the probe D-brane embedding in an acceleration black hole space time in Section 2. In the next Section 3, we move on towards discussing the thermodynamics in the boundary QFT at low temperature and finite density of  $U(1)$ charge, where the effects due to thermally produced charge pairs in the medium can be ignored. The main result of the paper is contained in Sections 4 and 5, where we discuss the Ohmic charge transport in the medium and its temperature dependencies considering different limiting situations. The scaling of the resistivity (in the boundary QFT) with the temperature motivates us to conjecture about the dynamic critical exponent ($z$) of the dual QFT to be equal to $3$. 

Finally, we conclude our discussion in Section 6.

\section{Accelerating black holes and DBI action}
We consider accelerating black holes in four dimensional AdS with metric \cite{Podolsky:2002nk}-\cite{Anabalon:2018ydc}
\begin{align}
ds^2_{Abh} = \Omega^{-2}(r , \theta)d\tilde{s}^2 
\end{align}
where $  \Omega(r , \theta)=(1+ A r \cos \theta)$ denotes the scale factor together with the line element 
\begin{align}
\label{e2.2}
d\tilde{s}^2  = -\frac{f(r)}{\alpha^2}dt^2 +\frac{dr^2}{f(r)}+r^2\Big( \frac{d\theta^2}{\Sigma}+\Sigma \sin^2\theta \frac{d\phi^2}{K^2} \Big).
\end{align}

The above entities in \eqref{e2.2} could be formally expressed as
\begin{align}
f(r)=(1- A^2 r^2)\Big( 1- \frac{2m}{r}\Big)+\frac{r^2}{\ell^2}~;~\Sigma (\theta)= 1+2m A \cos\theta
\end{align}
where $ A $ denotes the acceleration of the black hole and $ K $ is related to the deficit angle associated with the conical deficit\footnote{We choose $ K=1+2mA $, which leaves the conical deficit on the south pole of the two sphere \cite{Gregory:2017ogk}.}. Finally, here $ m = \alpha^{-1} K M $ is related to the mass ($ M $) of the black hole \cite{Appels:2018jcs}, together with the fact that the global time of $ AdS_4 $ is rescaled by a factor $ \alpha = \sqrt{1-A^2 \ell^2} $. As mentioned above, we would consider slowly accelerating ($ A \ell \ll 1 $) black holes in AdS. In this limit, one is left only with the black hole horizon and gets rid of the acceleration horizon.

The black hole horizon is given by solving the equation $ f(r)|_{r=r_+}=0 $, which yields three roots out of which only one is real and we denote it by $ r=r_+ $. The corresponding Hawking temperature can be obtained following the identity
\begin{align}
\label{e2.4}
\delta f(r)|_{r=r_+}=0=f'(r_+)\delta r_+ -\frac{2}{r_+}(1-A^2 r^2_+)\delta m - \frac{2 r^2_+}{\ell^3}\delta \ell - 2 A r^2_+ \Big(1-\frac{2m}{r_+} \Big)\delta A.
\end{align}

Using \eqref{e2.4}, we finally obtain the Hawking temperature of the black hole\footnote{In what follows, in the subsequent analysis, we set the AdS length scale $\ell =1$.}
 \cite{Anabalon:2018ydc}
\begin{align}
\label{e2.5}
T = \frac{f'(r_+)}{4 \pi \alpha}=\frac{1+3r^2_+ -A^2r^2_+(2+r^2_+ -A^2 r^2_+)}{4 \pi \alpha r_+ (1-A^2 r^2_+)}.
\end{align}

The boundary of the spacetime ($ r=r_b $), on the other hand, is given by the zeros of the scale factor $ \Omega $. This changes the usual boundary of AdS from $ r_b=\infty $ to $ r_b = - \frac{1}{A \cos\theta}$. Clearly, for $ \theta=\pi $, the boundary $ r_b(=A^{-1}) $ is closer than the usual boundary in AdS.

The purpose of this paper is to understand properties of the QFT dual to \eqref{e2.2}. In particular, we would be interested in computing various metallic transport coefficients associated with the dual QFT at finite density/chemical potential. Following \cite{Hartnoll:2009ns}, we probe the accelerating black hole spacetime \eqref{e2.2} using D-brane, that sources massless charge carriers for the dual QFT at strong coupling. Given an $ n $- dimensional manifold, the corresponding DBI action could be schematically expressed as
\begin{align}
\label{e2.6}
-\mathcal{S}_{DBI} = T_D \int d^n \xi \sqrt{- \det (g_{AB}+2\pi \alpha' \mathcal{F}_{AB})}+\mathcal{S}_{WZ}
\end{align}
where we ignore the WZ contribution for the time being. Here, $ g_{AB} $ is the induced world-volume metric, $ \mathcal{F}_{AB}=\partial_{[A} \mathcal{\mathcal{A}}_{B]} $ is the $ U(1) $ field strength tensor on the world-volume of D-brane and $ T_D $ represents the tension of the D-brane. The DBI \eqref{e2.6} sources massless charge carriers for the dual QFT, that are charged under the world-volume $ U(1) $. These charge excitations give rise to various thermodynamic properties as well as transport phenomena (for example, DC and Hall conductivities) that we discuss next.

To proceed further, we consider a D-brane living in the $ \theta = \theta_0$ subspace of the four dimensional spacetime \eqref{e2.2}, where one of the world-volume directions is stretched between the horizon ($ r_+ $) and the boundary ($ r_b)$. The  corresponding DBI Lagrangian reads as
\begin{align}
& - \det (g_{AB}+2\pi \alpha' \mathcal{F}_{AB})=\frac{ r^2 \Sigma  \sin ^2\theta_0}{\Omega ^6\alpha ^2 K^2}+\frac{4 \pi^2 \alpha'^2}{f \alpha^2 K^2 \Omega ^2}\Big(f^2 K^2 \mathcal{F}_{r \phi}^2 \nonumber\\&-\alpha ^2 \left(f r^2 \Sigma  \sin ^2\theta_0 \mathcal{F}_{tr}^2+K^2 \mathcal{F}_{t \phi}^2\right)\Big)
\end{align}
where, we keep $ \theta_0 $ to be arbitrary throughout our analysis.
\section{Thermodynamics at finite density}
Given the above set up, we now move on towards studying the thermodynamics of the dual QFT at finite chemical potential ($ \mu $). Here, we have perturbations in the form of an external driving force due to the cosmic string which slowly accelerates ($ A\ll 1 $) the black hole. This turns out to be the regime that allows us to treat this external force as a perturbation and write down a partition function considering a quasi-static equilibrium. In the language of the boundary QFT, the (grand-canonical) partition function could be expressed as a sum of the equilibrium partition function plus additional contributions proportional to the external driving agency, similar to that of a linear response theory.

The carriers are massless and are sourced due to probe D-brane \cite{Hartnoll:2009ns}-\cite{Lee:2010uy}. We turn on world-volume field $ \mathcal{A}_t(r) $, which sources a $ U(1) $ chemical potential ($ \mu $) for the dual QFT living on the boundary ($ r_b $). The world-volume theory reads as
\begin{align}
- \mathcal{S}_{DBI} = 2 \pi t_0 T_D \int_{r_+}^{r_b}dr \frac{\sqrt{\Sigma}r\sin\theta_0}{\alpha K \Omega^3}  \sqrt{1-\alpha^2 \Omega^4 \mathcal{A}_t'^2}+\mathcal{S}_{ct}
\end{align}
where we have absorbed the factor of $ 2\pi \alpha' $ into the definition of $ \mathcal{A}_t (r) $. Moreover, here $ 2\pi t_0 $ is an artefact of the $ \phi $ and time ($ 0\leq t \leq t_0$) integration. Here, the counter term can be schematically expressed as $\mathcal{S}_{ct} \sim \frac{1}{\varepsilon^2}\int dt d\phi$, whose detail will be given later on.

The grand-canonical free energy associated with the dual QFT is defined as
\begin{align}
\label{e3.2}
\mathcal{G}_b= - \mathcal{S}^{(on-shell)}_{DBI}+\mathcal{S}_{ct}.
\end{align}

The on-shell condition can be satisfied by solving the dynamics of the world-volume gauge field $\mathcal{A}_t$, which for the present case yields
\begin{align}
\label{e3.3}
\mathcal{A}_t'=\frac{1}{\alpha \Omega^2}\frac{1}{\sqrt{1+\frac{\beta^2 \Sigma \sin^2\theta_0 r^2}{\Omega^2 C^2 K^2}}}
\end{align}
where we denote $ \beta=2 \pi T_D t_0 $ and $C$ is the constant of integration\footnote{The constant $ C $ can indeed be related to the total charge ($ \mathcal{Q} $) of the D-brane, which is simply the flux of the world-volume two form $\mathcal{F}_2 = \mathcal{A}_t' dr \wedge dt$ across the co-dimension two surface of the D-brane. A straightforward computation reveals $ \mathcal{Q}=\frac{C K t_0 }{\alpha  \beta  h}\sinh ^{-1}\left(\frac{\beta  h r}{A C K r \cos \theta_0+C K}\right)\Big|_{r_+}^{r_b} $, where $ h=\sqrt{\Sigma}\sin\theta_0 $.}.

Using \eqref{e3.3}, the grand-canonical free energy \eqref{e3.2} turns out to be
\begin{align}
\label{e3.4}
\mathcal{G}_b = \beta^2 \int_{r_+}^{r_b}\frac{\Sigma \sin^2\theta_0 r^2}{\Omega^4 C K^2 \alpha}\frac{dr}{\sqrt{1+\frac{\beta^2 \Sigma \sin^2\theta_0 r^2}{\Omega^2 C^2 K^2}}}+\mathcal{S}_{ct}.
\end{align}

After performing the radial integration, one finds
\begin{align}
\label{e3.5}
\mathcal{ G}_b +  \frac{C}{2} \mu_b =\frac{C r}{ 2 \alpha}  \frac{ \sqrt{1+\frac{\beta ^2 \Sigma \sin^2\theta_0 r^2}{C^2 K^2 (1+A r \cos \theta_0)^2}}}{1+A r \cos \theta_0}\Big|_{r_+}^{r_b}+\mathcal{S}_{ct}
\end{align}
where $\mu_b = \int_{r_+}^{r_b}\mathcal{A}_t' dr$ is the chemical potential associated with the boundary QFT. Clearly, the L.H.S. of \eqref{e3.5} diverges near the boundary $ r\sim r_b $. This stems from the fact that the R.H.S. of \eqref{e3.5} diverges when expanded near the boundary $ r \sim r_b=-\frac{1}{A \cos\theta_0} $. To tame this divergence, one therefore needs to add a counter term $ \mathcal{S}_{ct}=\frac{\beta h}{2 \alpha K A^2 \varepsilon^2\cos^2\theta_0}$, where $ \varepsilon $ is the appropriate UV cut-off. In the presence of the suitable counter term, one can therefore express the regularised thermodynamic variables satisfying the relation
\begin{align}
\label{e3.6}
\bar{\mathcal{G}}_b +\frac{C}{2}\bar{\mu}_b=-\frac{C r_+}{ 2 \alpha}  \frac{ \sqrt{1+\frac{\beta ^2 \Sigma \sin^2\theta_0 r_+^2}{C^2 K^2 (1+A r_+ \cos \theta_0)^2}}}{1+A r_+ \cos \theta_0}
\end{align}
where bar denotes regularised entities.

Upon inverting \eqref{e2.5} and considering a slow acceleration of the black hole, one finds
\begin{align}
\label{e3.9}
r_+ = \frac{4 \pi  \alpha  T}{3}+\frac{2 A }{3 \sqrt{3} }\left(1+4 \pi ^2 \alpha ^2 T^2\right)+\mathcal{O}(A^2 T).
\end{align}

Given \eqref{e3.9}, we are now in a position where we can express the thermodynamic potential as a function of chemical potential and temperature namely $\bar{\mathcal{G}}_b = \bar{\mathcal{G}}_b  (\bar{\mu}_b , T)$, which is the standard way of expressing the Gibbs free-energy in a grand-canonical framework. We take \eqref{e3.6} as the defining relation for all our subsequent analysis, where we estimate various thermodynamic entities at low temperature $T\ll 1$ and finite density ($\bar{\mu}_b \neq 0$).

We first note down the charge density associated with the boundary QFT
\begin{align}
\bar{\varrho}_b = -\Big(\frac{\partial \bar{\mathcal{G}}_b}{\partial \bar{\mu}_b}\Big)_{T}=\frac{C}{2}.
\end{align}

Next, we compute the entropy in the grand-canonical ensemble
\begin{align}
\bar{s}_b = -\Big(\frac{\partial \bar{\mathcal{G}}_b}{\partial T}\Big)_{\bar{\mu}_b}=s_0+\frac{8 \pi ^2 \alpha  A T}{27 C K^2}\left(3 C^2 K^2 \left(\sqrt{3}-2 \cos \theta_0 \right)+2 \sqrt{3} \beta ^2 \Sigma \sin^2\theta_0\right)+\mathcal{O}(T^2)
\end{align}
where $ s_0=\frac{2 \pi  C}{3}$ is a constant, which corresponds to entropy at zero temperature.

The heat capacity, on the other hand, reveals an expression of the form
\begin{align}
\bar{c}_b=T\Big(\frac{\partial \bar{s}_b}{\partial T}\Big)_{\bar{\varrho}_b} = \frac{8 \pi ^2 \alpha  A T}{27 C K^2}\left(3 C^2 K^2 \left(\sqrt{3}-2 \cos \theta_0 \right)+2 \sqrt{3} \beta ^2 \Sigma \sin^2\theta_0\right)+\mathcal{O}(T^2)
\end{align}
which goes linear in $ T $, indicating a Fermi liquid behaviour \cite{Karch:2008fa} for this new quantum phase of matter, that is holographic dual to accelerating black holes in $AdS_4$ \cite{Karch:2009zz}-\cite{Lee:2010uy}.

Next, we note down the energy ($\bar{\epsilon}_b$) and pressure ($ \bar{p}_b $) for the boundary QFT
\begin{align}
\label{e3.13}
 \bar{\epsilon}_b= \bar{\mathcal{G}}_b+ \frac{C}{2} \bar{\mu}_b ~;~ \bar{p}_b =-\bar{\mathcal{G}}_b.
\end{align}

A straightforward calculation reveals the following expressions at low temperatures
\begin{align}
\label{e3.12}
& \bar{p}_b=\bar{p}_0 + \frac{16 \pi T A^2 \beta ^2 \Sigma \sin^2\theta_0}{81 C K^2}+\mathcal{O}(T^2)\\
&\bar{\epsilon}_b=\bar{\epsilon}_0+\frac{2 \pi  CT}{3}+\mathcal{O}(T^2).
\label{e3.13}
\end{align}
where, we denote the leading terms in the above expansions
\begin{align}
\bar{p}_0= \frac{8 A^3 \beta ^2 \Sigma \sin^2\theta_0}{243 \sqrt{3} \alpha  C K^2} ~;~\bar{\epsilon}_0=\frac{C A}{3 \sqrt{3} \alpha }
\end{align}
as respectively the pressure and energy associated with the boundary QFT at zero temperature. Clearly, at zero temperature and finite chemical potential, the pressure and the energy density of the dual QFT can be schematically expressed as $\bar{p}_0 \sim \bar{\mu}_0^3$ and $\bar{\epsilon}_0 \sim \bar{\mu}_0$, where $\bar{\mu}_0 \sim A$ is the associated chemical potential at zero temperature.

Using \eqref{e3.12} and \eqref{e3.13}, the enhancement in the speed of sound turns out to be
\begin{align}
\Delta v^2_s=\frac{8 A^2 \beta ^2 \Sigma \sin^2\theta_0}{81 C^2 K^2}.
\end{align}
\section{DC conductivity}
We now move on towards computing the DC conductivity in the dual QFT, following the methods of \cite{Karch:2007pd}. This approach is based upon first turning on the world-volume electric field $ \mathcal{E}=-\mathcal{F}_{t\phi}$ and thereby computing the current $ \mathcal{J}^\phi_b \sim \sigma_b \mathcal{E} $, where $ \sigma_b $ is the Ohmic conductivity in the boundary QFT. The first step is to define a new coordinate \cite{Lee:2010uy}
\begin{align}
\upsilon = \frac{1}{r}
\end{align}
which reveals the following 3d metric on the $ \theta= $ constant hyper-plane
\begin{align}
ds^2_{3d}=\Omega^{-2}(\upsilon, \theta_0)d\tilde{s}^2_{3d}
\end{align}
where $ \Omega (\upsilon, \theta_0)=\upsilon + A \cos\theta_0 $ and the new conformal boundary is located at $ \upsilon_{b}=-A \cos\theta_0 $.

The conformal 3d spacetime has the metric of the form
\begin{align}
d\tilde{s}^2_{3d}=-\frac{f(\upsilon)}{\alpha^2}dt^2+\frac{d \upsilon^{2}}{f(\upsilon)}+\Sigma\sin^2\theta_0 \frac{d\phi^2}{K^2}
\end{align}
where we list down the functions below
\begin{align}
\label{e26}
f(\upsilon)=1+(\upsilon^{2}-A^2)(1-2m \upsilon)~;~\Sigma = 1+2m A \cos\theta_0
\end{align}
in units where we set, $ \ell =1 $.

We propose the following ansatz for the world-volume gauge field
\begin{align}
\mathcal{A}=\mathcal{A}_t (\upsilon)dt + \mathcal{A}_\phi (t, \upsilon)d\phi~;~\mathcal{A}_\phi (t, \upsilon)=-\mathcal{E}t + \mathfrak{h}(\upsilon)
\end{align}
such that the corresponding field strength curvatures can be expressed as
\begin{align}
\label{e4.6}
\mathcal{F}_{t \upsilon}=-\mathcal{A}'_t (\upsilon)~;~\mathcal{F}_{t \phi}=- \mathcal{E}~;~\mathcal{F}_{\upsilon \phi}=\mathfrak{h}'(\upsilon).
\end{align}

Using \eqref{e4.6}, the DBI \eqref{e2.6} finally turns out to be
\begin{align}
\label{e4.7}
-\mathcal{S}_{DBI}=\beta \int_{\upsilon_+}^{\upsilon_b} d \upsilon \mathcal{L}_{DBI}
\end{align}
where the corresponding Lagrangian density reads as
\begin{align}
\label{e4.8}
\mathcal{L}_{DBI}=\sqrt{\frac{ \Omega ^4  \left(f^2 K^2 \mathfrak{h}'^2-\alpha ^2 \left(f \Sigma  \mathcal{A}_t'^2 \sin ^2\theta_0+K^2 \mathcal{E}^2\right)\right)+f \Sigma  \sin ^2\theta_0}{\alpha ^2 f K^2 \Omega ^6}}
\end{align}
together with the fact that we set, $ 2 \pi \alpha'=1 $  in natural units.

The equations of motion of $ \mathfrak{h} $ and $ \mathcal{A}_t $ lead to the following conserved entities
\begin{align}
\label{e4.9}
&\mathcal{H}=\frac{f^{3/2}K \Omega\mathfrak{h}'}{\alpha\sqrt{\Omega ^4  \left(f^2 K^2 \mathfrak{h}'^2-\alpha ^2 \left(f \Sigma  \mathcal{A}_t'^2 \sin ^2\theta_0+K^2 \mathcal{E}^2\right)\right)+f \Sigma  \sin ^2\theta_0}}\\
&\mathcal{C}=\frac{\alpha \sqrt{f} \Omega \Sigma \sin^2\theta_0 \mathcal{A}'_t}{K\sqrt{\Omega ^4  \left(f^2 K^2 \mathfrak{h}'^2-\alpha ^2 \left(f \Sigma  \mathcal{A}_t'^2 \sin ^2\theta_0+K^2 \mathcal{E}^2\right)\right)+f \Sigma  \sin ^2\theta_0}}.
\label{e4.10}
\end{align}

Using \eqref{e4.9}-\eqref{e4.10}, it is trivial find 
\begin{align}
\label{e4.11}
\mathcal{C}\mathfrak{h}' K^2 f = \alpha^2 \mathcal{H}\Sigma \sin^2\theta_0 \mathcal{A}_t'
\end{align}
which can be further used to solve
\begin{align}
\label{e4.12}
\mathcal{A}_t'^2=\frac{\mathcal{C}^2 K^2 \csc ^2\theta_0 \left(\alpha ^2 K^2 \Omega ^4 \mathcal{E}^2 \csc ^2\theta_0-f \Sigma \right)}{\alpha ^2 \Sigma  \Omega ^2 \left(\mathcal{C}^2 f K^2 \Omega ^2 \csc ^2\theta_0+\Sigma  \left(f-\alpha ^2 \Omega ^2 \mathcal{H}^2\right)\right)}.
\end{align}

Upon integrating \eqref{e4.12}, one finds an expansion near the boundary $\upsilon \sim \upsilon_b \sim 0$
\begin{align}
\mathcal{A}_t(\upsilon \sim 0)=\mu_b - \frac{ \mathcal{C} K }{\alpha}\log \upsilon + \mathcal{O}(\upsilon^2)
\end{align}
which reveals that the charge density for the boundary theory, $ \mathcal{J}^t_b = \frac{ \mathcal{C} K }{\alpha} $.

On a similar note, one finds
\begin{align}
\mathfrak{h}(\upsilon \sim 0)=\mathfrak{h}_0 +\frac{\alpha  \mathcal{H}}{\left(1-A^2\right) K }\log\upsilon+\mathcal{O}(\upsilon)
\end{align}
where $ \mathfrak{h}_0  $ is a constant together with the boundary current density, $ \mathcal{J}^\phi_b =  \frac{\alpha  \mathcal{H}}{\left(1-A^2\right) K }$.

Using \eqref{e4.11}-\eqref{e4.12}, the DBI Lagrangian \eqref{e4.8} finally turns out to be
\begin{align}
\label{e4.15}
\mathcal{L}_{DBI}=\sqrt{\frac{\left(f-\alpha ^2 K^2 \upsilon ^4 \mathcal{E}^2\right) \left(2 \upsilon ^2 \left(\mathcal{C}^2 f K^2-\alpha ^2 \mathcal{H}^2\right)+f\right)}{\alpha ^2 f K^2 \upsilon^6 \left(\upsilon ^2 \left(\mathcal{C}^2 f K^2-\alpha ^2 \mathcal{H}^2\right)+f\right)}}\equiv \sqrt{\frac{\mathcal{N}}{\mathcal{D}}}
\end{align}
where we set $ \theta_0 = \frac{\pi}{2} $ which sets the boundary $ \upsilon_b \sim 0 $.

At this stage, we follow the prescription due to \cite{Karch:2007pd}, which is to ensure that \eqref{e4.15} remains positive definite throughout the range $ \upsilon \sim \upsilon_+ $ to $ \upsilon \sim  \upsilon_b \sim 0 $. An expansion near the horizon reveals the following functions
\begin{align}
\mathcal{N}(\upsilon \sim \upsilon_+)=2 \alpha ^4 K^2 \upsilon^6_+ \mathcal{E}^2 \mathcal{H}^2 ~;~ \mathcal{D}(\upsilon \sim \upsilon_+)=- \alpha^4 \upsilon^8_+ K^2 \mathcal{H}^2 (\upsilon - \upsilon_+)f'(\upsilon_+)
\end{align}
both of which are positive definite subjected to the fact that $ f'(\upsilon_+) <0 $. Clearly, one has to add an appropriate IR regulator when evaluating the integral \eqref{e4.7} near the horizon. 

On a similar note, a near boundary expansion reveals
\begin{align}
\mathcal{N}(\upsilon \sim 0)=(1-A)^2 ~;~\mathcal{D}(\upsilon \sim 0)=\alpha ^2 K^2 \left(1-A^2\right)^2  \upsilon ^6
\end{align}
that ensures to include an appropriate UV regulator while evaluating \eqref{e4.7} near UV.

To summarise, the ratio $ \frac{\mathcal{N}}{\mathcal{D}} $ suffers from positive divergences near the horizon ($ \upsilon_+ $) and the asymptotic boundary ($ \upsilon_b $). Therefore, the function must attain a minima at $\upsilon_\ast$ where $\upsilon_b< \upsilon_\ast<\upsilon_+ $. This yields the following set of conditions
\begin{align}
\label{e40}
&\mathcal{N}\Big|_{\upsilon = \upsilon_\ast}=\mathcal{N}_{\ast}=\left(f_\ast-\alpha ^2 K^2 \upsilon_\ast ^4 \mathcal{E}^2\right) \left(2 \upsilon_\ast ^2 \left(\mathcal{C}^2 f_\ast K^2-\alpha ^2 \mathcal{H}^2\right)+f_\ast\right)\\
&\mathcal{D}'\Big|_{\upsilon = \upsilon_\ast}=0=f_\ast^2 \left(8 \mathcal{C}^2 K^2 \upsilon_\ast ^2+6\right)+2 \upsilon_\ast  f_\ast \left(f'_\ast \left(\mathcal{C}^2 K^2 \upsilon_\ast ^2+1\right)-4 \alpha ^2 \upsilon_\ast  \mathcal{H}^2\right)-\alpha ^2 \upsilon_\ast ^3 \mathcal{H}^2 f'_\ast
\label{e41}
\end{align}
where $ \mathcal{D}(\upsilon) $ has a maxima at $ \upsilon = \upsilon_\ast $ together with $ \mathcal{N}_{\ast}\geq 0 $ and $ f_\ast = f(\upsilon_\ast) $.

To solve \eqref{e40}, we choose $ \mathcal{N}_\ast =0 $, which corresponds to shifting the origin such that the minima of the function $  \frac{\mathcal{N}}{\mathcal{D}}\Big|_{\upsilon = \upsilon_\ast}=0 $. It is now trivial to find the solution
\begin{align}
\label{e42}
f_\ast = \alpha ^2 K^2 \upsilon_\ast ^4 \mathcal{E}^2.
\end{align}

The above equation \eqref{e42} can be solved perturbatively 
\begin{align}
\upsilon_\ast = \upsilon_\ast^{(0)}+\mathcal{E}^2 \upsilon_\ast^{(1)}+\mathcal{O}(\mathcal{E}^4)
\end{align}
where the electric field $ \mathcal{E}(\ll 1) $ is treated as an expansion parameter.

The corresponding solutions may be listed down as
\begin{align}
\upsilon_\ast^{(0)}=\frac{1}{6 m}-\mathfrak{V}(m,A)~;~\upsilon_\ast^{(1)}=\frac{\alpha ^2 K^2 (1-6 m \mathfrak{V}(m,A))^4}{216 \left(12 m^5 \left(A^2-3 \mathfrak{V}(m,A)^2\right)+m^3\right)}
\end{align}
where the function $ \mathfrak{V}(m,A) $ is given below
\begin{align}
& \mathfrak{V}(m,A) =\frac{12 A^2 m^2+1}{3\ 2^{2/3} m \sqrt[3]{72 A^2 m^2+\sqrt{\left(36 \left(3-2 A^2\right) m^2+2\right)^2-4 \left(12 A^2 m^2+1\right)^3}-108 m^2-2}}\nonumber\\
& +\frac{\sqrt[3]{72 A^2 m^2+\sqrt{\left(36 \left(3-2 A^2\right) m^2+2\right)^2-4 \left(12 A^2 m^2+1\right)^3}-108 m^2-2}}{6 \sqrt[3]{2} m}.
\end{align}

On the other hand, from \eqref{e41} we find
\begin{align}
\label{e46}
\mathcal{H}^2=\frac{2 f_\ast \left(\upsilon_\ast  f'_\ast \left(\mathcal{C}^2 K^2 \upsilon_\ast ^2+1\right)+f_\ast\left(4 \mathcal{C}^2 K^2 \upsilon_\ast ^2+3\right)\right)}{\alpha ^2 \upsilon_\ast ^2 \left(\upsilon_\ast  f'_\ast+8 f_\ast\right)}.
\end{align}

Using \eqref{e26} and \eqref{e46} and considering the fact that\footnote{This is a reasonable assumption in the limit when $A \ll 1$ and the black hole mass ($m$) is large enough such that their product $m A^2$ is always finite and less than unity.} $\upsilon_\ast \ll 1$,  finally reveals\footnote{We have absorbed the factor of $\alpha^{-1} (1-A^2)K $ into the definition of $ \mathcal{J}^\phi_b $.}
\begin{align}
&(\mathcal{J}^\phi_b)^2 =\mathcal{J}^{2}_0+\sigma_b^2 \mathcal{E}^2 \\
&\mathcal{J}^{2}_0 = (\mathcal{J}^t_b)^2 +\frac{\upsilon_\ast^{2(0)}(17-39 m \upsilon_\ast^{(0)})+12}{16 \alpha ^2 \upsilon_\ast^{2(0)}}-\frac{A^2}{32 \alpha^2 \upsilon^{2(0)}_\ast}\Big(8 \left(4 \alpha ^2 (\mathcal{J}^t_b)^2 \upsilon^{2(0)}_\ast+3\right)\nonumber\\&- m \upsilon^{(0)}_\ast \left(\upsilon^{2(0)}_\ast \left(72 \alpha ^2 (\mathcal{J}^t_b)^2-5\right)+58\right)\Big)
\label{e48}
\end{align}
where $\mathcal{J}_0 (m , A)$ is a steady background current that appears purely due to the acceleration of the D-brane. This is the physical acceleration of the D-brane that produces an acceleration of the (thermally produced as well as $ U(1) $) charge carriers of the medium, and thereby causing steady current contribution even in the absence of an external electric field. The dependence is not explicit on the acceleration ($ A $), however it has its root at the very beginning of the calculation where we set $ \theta =\theta_0$ in order to identify a particular location of the boundary point ($ \upsilon_b $). Setting a particular $ \theta $ has a deep connection with the acceleration of the D-brane since different orientation would correspond to different location of the boundary with respect to the horizon ($\upsilon_+$) of the accelerating black hole. 

Finally, the Ohmic conductivity in the boundary QFT appears to be
\begin{align}
\label{e49}
&\sigma_b/\sigma_0=  \Big[8 m \left(\frac{A^2 \left(72 \alpha ^2 \left(1-A^2\right) (\mathcal{J}^t_b)^2+73\right)-78}{1-A^2}-\frac{58 A^2}{\upsilon_\ast^{2(0)}} \right)\nonumber\\
&-\frac{384 \left(1-A^2\right)}{\upsilon_\ast^{3(0)}}+\frac{45 A^6 m^3}{\left(1-A^2\right)^2}\Big]^{1/2}
\end{align}
where $ \sigma_0 =\frac{\sqrt{\upsilon^{(1)}_\ast}}{16 \alpha} $ is an overall scaling factor.

The term proportional to $\mathcal{J}^t_b$ reflects the direct contribution of the $U(1)$ charge carriers to the boundary conductivity. These charge carriers are sourced due to the probe D-brane added to the bulk. On the other hand, we have pure thermal contribution to the boundary conductivity those are proportional to (inverse) powers of $|\upsilon^{(0)}_\ast| \sim \zeta m^{-1/3}$. The parameter $m$ is (equivalent to the mass of the black hole \cite{Appels:2016uha}, \cite{Appels:2018jcs}) what we identify as thermal effects in the presence of accelerating black hole, which stems from the fact that in the limit $m \rightarrow 0$, one is left with pure AdS and with no thermal (Hawking) radiation \cite{Anabalon:2018ydc}.

Temperature ($T$) of the black hole can be expressed in terms of inverse horizon radius ($ \upsilon_+ $), mass ($ m $) and acceleration ($ A $), which reads as \cite{Appels:2016uha}, \cite{Appels:2018jcs}
\begin{align}
T=\frac{1}{2 \pi  \upsilon_+}(A^2 (m \upsilon_+ -1)+m \upsilon_+^3+1)
\end{align}
which can be inverted to obtain
\begin{align}
\label{e51}
m=\frac{2 \pi  T}{\upsilon_+^2}+\mathcal{O}(m A^2).
\end{align}

Using \eqref{e51}, we can express the background current \eqref{e48} and the Ohmic conductivity \eqref{e49} in the boundary QFT as a function of temperature ($ T $), which reads as 
\begin{align}
\label{e52}
&\mathcal{J}^{2}_0 = (\mathcal{J}^t_b)^2 +\Big(\frac{17}{16 \alpha^2} +\frac{3 (2 \pi )^{2/3}(4-13 \zeta ^3)}{16 \alpha^2 \zeta^2 \upsilon^{4/3}_+}T^{2/3}\Big)+\mathcal{O}(m^{2/3}A^2)\\
&\sigma_b/\sigma_0=\frac{2 \sqrt{2 \pi T}}{\upsilon^3_+}\Big[ 2 \upsilon_+^4 \left(A^2 \left(\frac{73}{1-A^2}+72 \alpha ^2 (\mathcal{J}^t_b)^2-\frac{58 (2 \pi )^{2/3}T^{2/3}}{\zeta ^2 \upsilon^{4/3}_+}\right)-\frac{78}{1-A^2}\right)\nonumber\\
&+ \frac{45 \pi ^2 A^6 T^2}{\left(1-A^2\right)^2} -\frac{96 \upsilon_+^4}{\zeta ^3}\left(1-A^2\right) \Big]^{1/2}.
\label{e53}
\end{align}
\section{Resistivity}
Given \eqref{e52}-\eqref{e53}, one can estimate resistivity in two different limits. In the first approximation, we consider the regime where $ U(1) $ charge current ($ \mathcal{J}^t_b $) dominates over the current that is produced due to thermally excited charge pairs. Clearly, this is the domain where the temperature ($ T $) is low enough to produce thermally excited charge pairs in the medium. This results into a steady current that can be schematically expressed as 
\begin{align}
(\mathcal{J}^\phi_b)^2 =(\mathcal{J}^t_b)^2+\mathfrak{R}_b^{-2} \mathcal{E}^2
\end{align}
where the Ohmic resistivity due to $ U(1) $ carriers turns out to be
\begin{align}
\mathfrak{R}_b = \frac{8 \pi^{2/3}T^{2/3}}{3 . 2^{1/6}\alpha K \upsilon_+^{4/3}\mathcal{J}^t_b} 
\end{align}
which exhibits a new phase of quantum matter \cite{Karch:2009zz}, where the resistivity scales differently (in the $ U(1) $ dominated phase) than the usual strange metallic systems \cite{Hartnoll:2009ns}. A closer comparison with \cite{Hartnoll:2009ns}, \cite{Lee:2010uy} reveals that the boundary QFT acts like a quantum critical point with dynamic exponent $ z=3 $. This further reflects to the fact that the bulk in the dual accelerating black hole phase does not possesses usual AdS asymptotic.

The second limiting situation arises when $ U(1) $ contribution becomes negligible as compared to the thermally generated current. This is the phase that corresponds to much higher temperature, where thermally excited charge pairs become populated over $ U(1) $ charged carriers added externally due to probe D-brane. A straightforward computation reveals that the leading order contribution in the boundary current is of the form 
\begin{align}
(\mathcal{J}^\phi_b)^2 =\frac{3 (2 \pi )^{2/3}(4-13 \zeta ^3)}{16 \alpha^2 \zeta^2 \upsilon^{4/3}_+}T^{2/3}+\mathfrak{R}_b^{-2}\mathcal{E}^2
\end{align}
where the back ground current due to acceleration scales as $T^{2/3}$. On the other hand, the Ohmic resistivity due to thermally generated charge pairs becomes 
\begin{align}
\mathfrak{R}_b=\frac{8. 2^{11/6}(1-A^2)\upsilon_+^{2/3}}{3.5^{1/2}\pi^{1/3}A^2}T^{-1/3}.
\end{align}

Clearly, the LO contribution to the Ohmic resistivity scales as $T^{-1/3}$. Considering a generic form (at zero $U(1)$ charge density) $T^{-|p-2|/z}$ for a $(p+2)$ dimensional space time, where $p$ is the number of spatial dimensions that the probe D-brane explores in the bulk, the dynamic critical exponent again turns out to be $z=3$.  Notice that, in the present calculation $p=1$, as the D-brane moves only in the subspace defined by $(t, r, \phi)$ coordinates, where $\phi$ plays the role of the spatial direction in our analysis.
\section{Conclusions and Outlook}
We now summarise the key findings of the paper. The purpose of the present paper is to explore charge transport phenomena due to accelerating black holes in four dimensions and in the presence of a negative cosmological constant. We work in the slow acceleration ($A \ll 1$) limit which turns out to be crucial in defining a equilibrium partition function for the dual QFT at strong coupling. The acceleration sources an external ``driving force'' for the boundary QFT, which can be treated as a perturbation over a quasi-equilibrium configuration. In other words, we can define an (quasi)equilibrium partition function and compute various response functions at strong coupling and finite density. 

The method employed is the celebrated gauge/gravity duality applied to finite density systems. The DC current in the boundary QFT receives contributions both due to the acceleration of the black hole (that in turn produces an acceleration of the D-brane world-volume) and the world-volume electric field ($ \mathcal{E} $) of the probe D-brane. The constant background current is what we identify as the new ingredient in the accelerating black hole story. Physically this means that the charge carriers in the world-volume move with a uniform acceleration ($ A \ll1 $) through the spacetime and thereby producing a constant current. This corresponds to an additional ``driving force'' on the charge carriers in the boundary QFT, other than the usual force due to the world-volume electric field ($ \mathcal{E} $). 

The DC charge current computed for the boundary QFT receives contributions due to - (i) $ U(1) $ charged carriers injected by the probe D-brane and (ii) thermally excited charge pairs in the presence of a black hole. As anticipated, thermally excited charge pairs become overpopulated as temperature increases above the ground state. In other words, thermally induced current takes over the $ U(1) $ charge current ($ \mathcal{J}_b^t $) at higher temperatures.

One could think of these charge carriers as being some form of excitations of a strongly coupled QFT that are drifted through the medium. While these carriers propagate through the medium, the medium offers a drag force that balances the external forces and thereby producing a steady current. This takes us to the immediate question of resistivity, particularly in the context of Ohmic transports. We estimate resistivity in two limiting situations. First we consider the situation when temperature is not large enough to produce thermally excited charge pairs, which makes the $ U(1) $ charge current ($ \mathcal{J}_b^t $) dominant over the thermally induced current. Our analysis reveals that the corresponding resistivity exhibits a different scaling, $ \mathfrak{R}_b \sim T^{2/3} $ than those expected for a $ z=2 $ fixed point  \cite{Hartnoll:2009ns}, \cite{Lee:2010uy}, which is therefore a new class of quantum liquid phase at finite density. a similar analysis, in the limit of negligible $U(1)$ charge density reveals $ \mathfrak{R}_b \sim T^{-1/3} $. Combining the above two features together, we conclude that accelerating black holes are dual to a strongly coupled QFT with dynamic critical exponent $z=3$.
\subsection*{Acknowledgments}
DR would like to acknowledge The Royal Society, UK for financial assistance. DR also acknowledges the Mathematical Research Impact Centric Support (MATRICS) grant (MTR/2023/000005) received from ANRF, India.

\end{document}